\documentclass[aps,apl,amsmath,amssymb,
reprint,
superscriptaddress
]{revtex4-2}

\usepackage{graphicx} 
\usepackage[margin=0.75in]{geometry} 

\usepackage{geometry}

\usepackage{todonotes}
\usepackage{xspace}
\usepackage{placeins}
\usepackage{mathtools} 
\usepackage{braket}
\usepackage{adjustbox}

\usepackage{hyperref}
\usepackage{nameref} 

\newcommand{\micro}{\textrm{$\mu$}}

\newcommand{\um}{$\mu$m}

\begin{document}

\title{Quality of Helicity-Dependent Magnetization Switching by Phonons}
\author{F.G.N. Fennema }
\email{niels.fennema@ru.nl}
\affiliation{FELIX Laboratory, Radboud University, Toernooiveld 7, 6525 ED Nijmegen, The Netherlands}

\author{C. S. Davies}
\affiliation{FELIX Laboratory, Radboud University, Toernooiveld 7, 6525 ED Nijmegen, The Netherlands}

\author{A. Tsukamoto}
\affiliation{College of Science and Technology, Nihon University, Chiba, Japan.}

\author{A. Kirilyuk}
\affiliation{FELIX Laboratory, Radboud University, Toernooiveld 7, 6525 ED Nijmegen, The Netherlands}

\date{\today}


\begin{abstract}
    Optical control of magnetization has emerged as a promising approach to achieve ultrafast and energy-efficient magnetization reversal. Here, we investigate helicity-dependent switching of magnetization driven by the resonant excitation of circularly-polarized transverse-optical phonons, using a polarization-modulated transient grating. Our results show that the polarized phonons within the sample substrate induce robust, helicity-defined magnetization reversal in the magnetic overlayer. Moreover, the quality of switching remains largely unaffected when the degree of ellipticity of the infrared excitation is varied at frequencies resonant with the targeted phonon modes. Conversely, as the excitation is moved slightly off-resonance, switching quality becomes highly sensitive to the ellipticity of the incident light. 
   
\end{abstract}

\maketitle

In recent decades, the exponential growth in global data collection has led to pessimistic projections of a three- to four-fold increase in data center energy consumption \cite{masanet_recalibrating_2020}. Despite technological advances, $90\%$ of all data recorded in data centers is still stored on hard drive disks (HDDs) \cite{seagate_technology_llc_why_nodate}, highlighting the continued dominance of magnetic storage. In HDDs, magnetic bits are written using a spinning disk and a static electromagnet, which together consume several watts. Therefore, identifying more energy-efficient and faster magnetization switching pathways is essential for sustainable data storage.

Advances in the research field of ultrafast magnetism have revealed a potentially more efficient switching method than that employed in present-day HDDs, based on the use of ultrashort optical pulses  \cite{stanciu_all-optical_2007,kirilyuk_laser-induced_2013}. This approach allows the magnetization in specific ferrimagnetic alloys to be all-optically toggled by the thermal load delivered by the incident pulse \cite{vahaplar_ultrafast_2009,radu_transient_2011,ostler_ultrafast_2012,mentink_ultrafast_2012,davies_helicity-independent_2022}. Follow-up experiments showed that this exchange-driven process can become helicity-dependent due to the differential optical absorption introduced by magnetic circular dichroism (MCD) \cite{khorsand_role_2012}. Later, a different helicity-dependent all-optical switching (HD-AOS) mechanism was identified in a ferromagnetic multilayer, driven by multiple pulses \cite{mangin_engineered_2014,lambert_all-optical_2014}. In this mechanism, initial laser pulses create a multidomain state through thermal demagnetization, with subsequent pulses inducing a helicity-dependent temperature gradient via MCD, driving domain wall motion. This domain-wall motion can deterministically result in domain expansion or erasure, leading to the emergence of helicity-dependent switching  \cite{medapalli_multiscale_2017,kichin_multiple-_2019,yamada_efficient_2022}.

Subsequent research revealed non-thermal approaches for magnetization switching in dielectric iron-garnets. One mechanism involves the excitation of specific electronic states based on the orientation of the optical pump polarization. These excitations selectively lift the degeneracy between magnetization states, inducing magnetic anisotropy that guides perturbed spins to settle into the energetically favored magnetic state  \cite{stupakiewicz_ultrafast_2017}. More recently, Stupakiewicz et al.\ demonstrated that switching in these garnets can also be achieved by driving particular optical phonons at resonance \cite{stupakiewicz_ultrafast_2021}. This latter mechanism is a promising avenue to achieve both non-thermal and universal means of magnetic switching.  However, a key drawback of this effect is that subsequent pulses appear to be unable to restore the magnetization to its original state.

Our recent work \cite{davies_phononic_2024} presented an alternative approach to magnetization switching via the resonant excitation of circularly polarized transverse-optical (TO) phonons, not within the ferrimagnetic material but rather in the paramagnetic substrate \cite{dornes_ultrafast_2019,tauchert_polarized_2022,juraschek_phono-magnetic_2020,juraschek_phono-magnetic_2020,basini_terahertz_2024,luo_large_2023}. Our measurements revealed that the quality of magnetization reversal in the magnetic overlayer followed the absorption spectra of the sapphire and glass-ceramic substrates. Using a single wavelength-tuned pulse resulted in only demagnetization, while multiple pulses yielded magnetization reversal. This suggests a multi-step process in which the initial magnetization of the magnetic layer must diminish before the circularly polarized optical phonons can successfully drive the switching process. By varying the helicity of the pump light, we can selectively excite left- or right-handed circularly polarized optical phonons, resulting in helicity-defined magnetization reversal. 

However, key aspects of the underlying mechanism remain unclear. In particular, the degree of circular polarization required to effectively excite the circular phonons at the given wavelength was unknown.  
Moreover, the achievable wavelengths were constrained to below 21 $\mu$m by the absorption properties of the CdSe-based quarter-waveplate. 

In this letter, an optical transient grating was employed to extend the wavelength and polarization range and thus further explore this mechanism. The transient grating offers the possibility to continuously vary the optical polarization across the sample, enabling the study of its influence on switching efficiency and the extension of the wavelength range beyond the quarter-waveplate's capabilities.
In this excitation scheme, two orthogonally-polarized infrared laser pulses intersect at an angle and interfere, generating a spatially-varying polarization with continuous intensity, providing an ideal environment to measure the helicity dependence of the mechanism. Moreover, this scheme is independent of the optical spectrum, allowing for rapid changes in excitation wavelength with limitations arising only from the light source itself. Our results reveal that when the pump wavelength is resonant with a targeted phonon, a high degree of circular polarization is not crucial. Conversely, when the light is slightly off-resonance, a small reduction in circularity significantly diminishes the quality of the switching.


The sample used in this investigation consisted of four layers. At the base is a 0.5-mm-thick double-side polished c-cut sapphire substrate commercially obtained from Alineason. Atop this substrate, magnetron sputtering was used to deposit a 10-nm-thick $\mathrm{Si}_{3}\mathrm{N}_{4}$ interlayer, a 20-nm-thick amorphous ferrimagnetic $\mathrm{Gd}_{24}\mathrm{Fe}\mathrm{Co}$ film and a 60-nm-thick $\mathrm{Si}_{3}\mathrm{N}_{4}$ capping layer \cite{stanciu_all-optical_2007,stanciu_ultrafast_2006}.

Infrared (IR) optical pulses from the FELIX facility in Nijmegen, the Netherlands, were used to excite circularly-polarized phonons at resonance. The free-electron laser (FEL) used is widely tunable, offering a wavelength range of $3$ to $ 120\ \mathrm{\micro} $m and a pulse bandwidth between 0.5 and 2$\%$ \cite{oepts_free-electron-laser_1995}. Due to the use of normal-conducting linacs, the light generated by the FEL is emitted in bursts (''macropulses'') with a typical length between $2$ and $10\ \mathrm{\micro} $s, at a frequency of 10 Hz. Each macropulse consists of shorter pulses (''micropulses'') with energies on the order of 10 $\mathrm{\micro}$J, pulse durations between 0.25 and 7 ps, and a tunable repetition rate between 25 MHz and 1 GHz.

\begin{figure}[h]
    \centering
    \includegraphics[width=\linewidth]{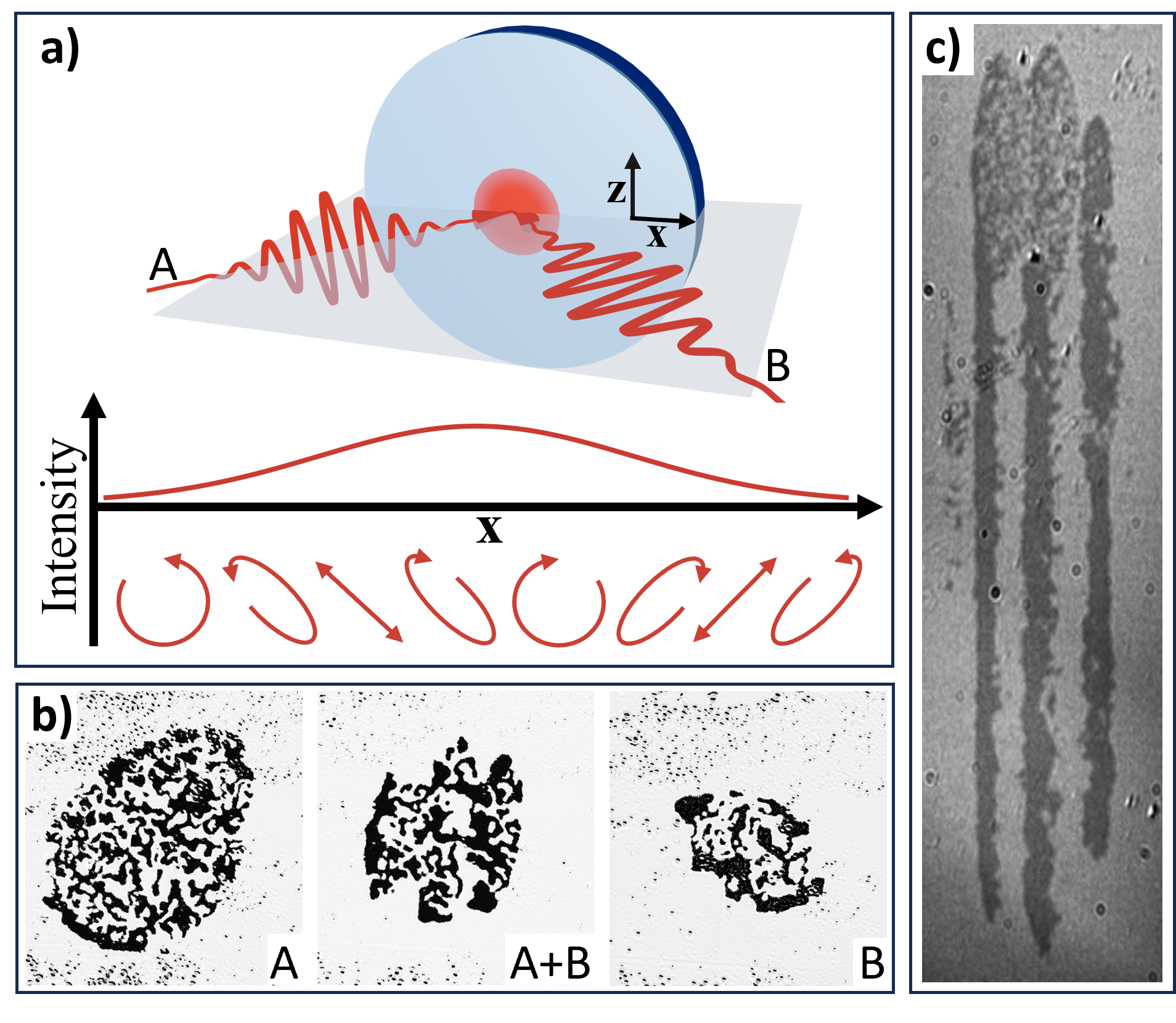}
    \caption{Schematic overview of transient grating. (a) Two light pulses (marked A and B) with orthogonal polarizations impinge on the sample, creating a grating with spatially varying polarization. (b) Magneto-optical images showing the effect of exposing the sample to pulse A and pulse B separately (left and right panels), and the combination of both (middle panel). Pure demagnetization results from the absorption of linearly polarized light originating from arms A and B, respectively. The combination of the light from arms A and B creates a spatially varying polarization that results in switching. (c) Magneto-optical image showing the effect of sweeping the transient grating ($\lambda$ = 17 \micro m) across the sample surface from bottom to top at 20 \micro m/s.}
    \label{fig:fig1}
    
\end{figure}

 To generate the polarization-modulated transient grating, linearly polarized IR pulses from the FEL were split by a pellicle beamsplitter into transmitted and reflected beams. The transmitted beam passed through a polarization rotator \cite{knippels_fel_1998}, rotating its polarization by 90$^{\circ}$, and was then focused on the sample via a 3-inch concave mirror with a 150 mm focal length. The reflected beam was sent across a retroreflector mounted on a motorized delay line to equalize the beam paths, and was then focused onto the sample surface by the same concave mirror, forming an angle of 29$^{\circ}$ between the overlapping beams. The interference between the two orthogonally polarized light pulses created a spot with spatially varying polarization and continuous intensity \cite{janusonis_magnetoelastic_2017,davies_towards_2018}, see Fig.\ \ref{fig:fig1}a, enabling simultaneous helicity-dependent measurements across the sample.  
  
 The effect of the IR light on the magnetization of the sample was probed several seconds after optical irradiation using a Faraday microscope equipped with an LE.5211 cold light source, two polarizers, a 20× objective lens, and a Thorlabs Quantalux scientific sCMOS camera. 

Fig. \ref{fig:fig1}b shows the effect of light pulses on the magnetic layer. The sizes of spots A and B vary due to the asymmetry in the transmittance and reflection of the beamsplitter. Since both spots are linearly polarized, they individually create a demagnetized multidomain magnetic state. A transient grating forms when the two pulses overlap. 
Due to the Gaussian profile of the light pulses, intensity peaks at the center of the grating, leading to elevated temperatures that leave the sample in a partially demagnetized state. Toward the edges, the temperature is sufficiently lower to allow the formation of larger magnetic domains, slightly revealing the interference pattern. To make this effect more evident, the sample is moved across the laser spot, leaving a track, as illustrated in Fig.\ \ref{fig:fig1}c.

\begin{figure}[htbp!]
    \includegraphics[width=0.7\linewidth]{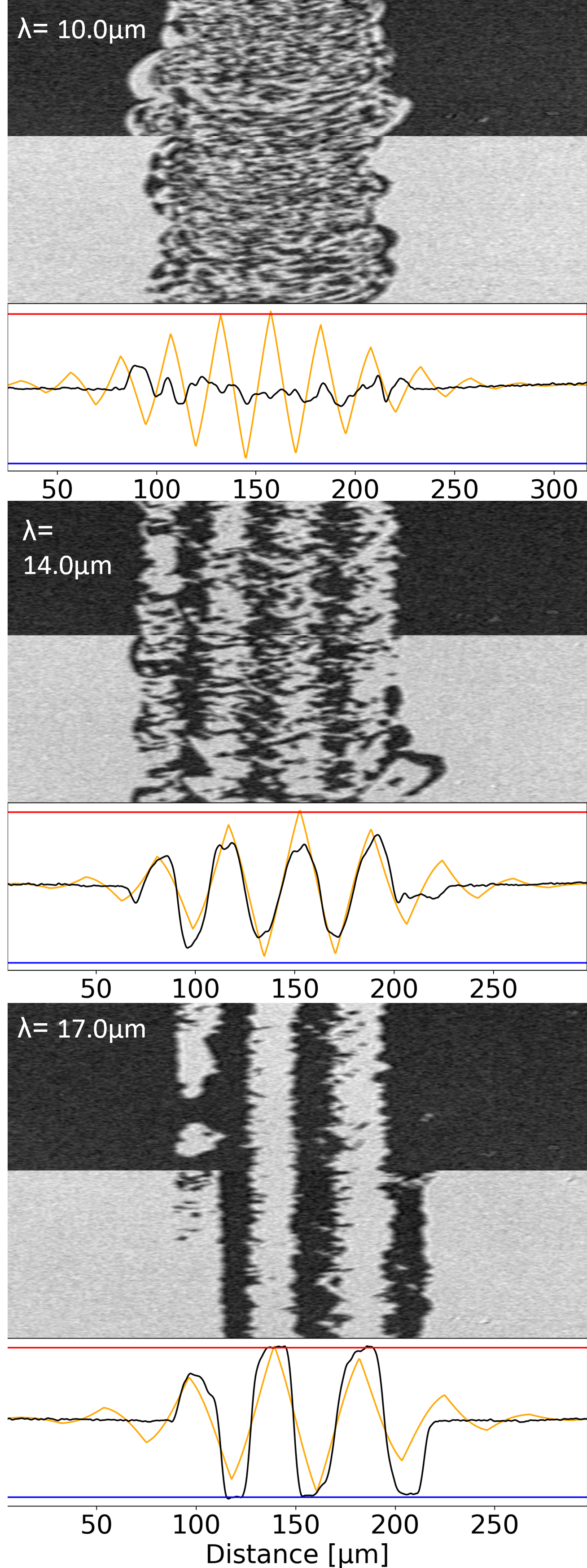}
    \caption{Spectral dependence of observed switching efficiency. For every pumping wavelength $\lambda$ shown, two images are stacked to evaluate the observed switching as a function of the background. Below the images, a black trace is shown to indicate the magnetization state of the corresponding pixel column. The orange trace shows the calculated ellipticity across the sample. With $\lambda~=~$10 \um\, no switching can be discerned, while for $\lambda~=~$14 \um\, switching becomes apparent and overlaps with the helicity trace. Finally, when $\lambda~=~$17 \um\, switching peaks and $100 \% $ switching efficiency is observed. The red and blue horizontal lines shown in the graphs are the values of a perfectly switched light or dark column, respectively.}
    \label{fig:fig2}
\end{figure}

To evaluate the quality of switching in these tracks, we scanned the same polarization pattern over the sample magnetized in opposite directions. Then, two cropped images with opposite magnetization backgrounds were stacked, as shown in Fig.\ \ref{fig:fig2}, and all the vertical pixels were summed. This summation is displayed as a trace below the images in Fig.\ \ref{fig:fig2} to indicate the magnetization state of each pixel column. For direct comparison, the calculated ellipticity from our transient grating, plotted as a function of distance across the sample, is overlaid in orange. The polarization is maximally circular at the extrema of the orange trace, with the maxima and minima corresponding to opposite helicities.

This helicity was calculated by passing two linearly polarized light pulses, angled at a variable angle, through left- and right-circular polarizers, resulting in overlapping circular light. This created a phase shift between the two pulses across the sample, which is also dependent on the light wavelength. By subtracting the amplitude of one helicity from the other, the net helicity was obtained and then plotted as the orange trace (see Supplementary Note 4: Jones Calculus for details).

\begin{figure}[h]
    \centering
    \includegraphics[width=\linewidth]{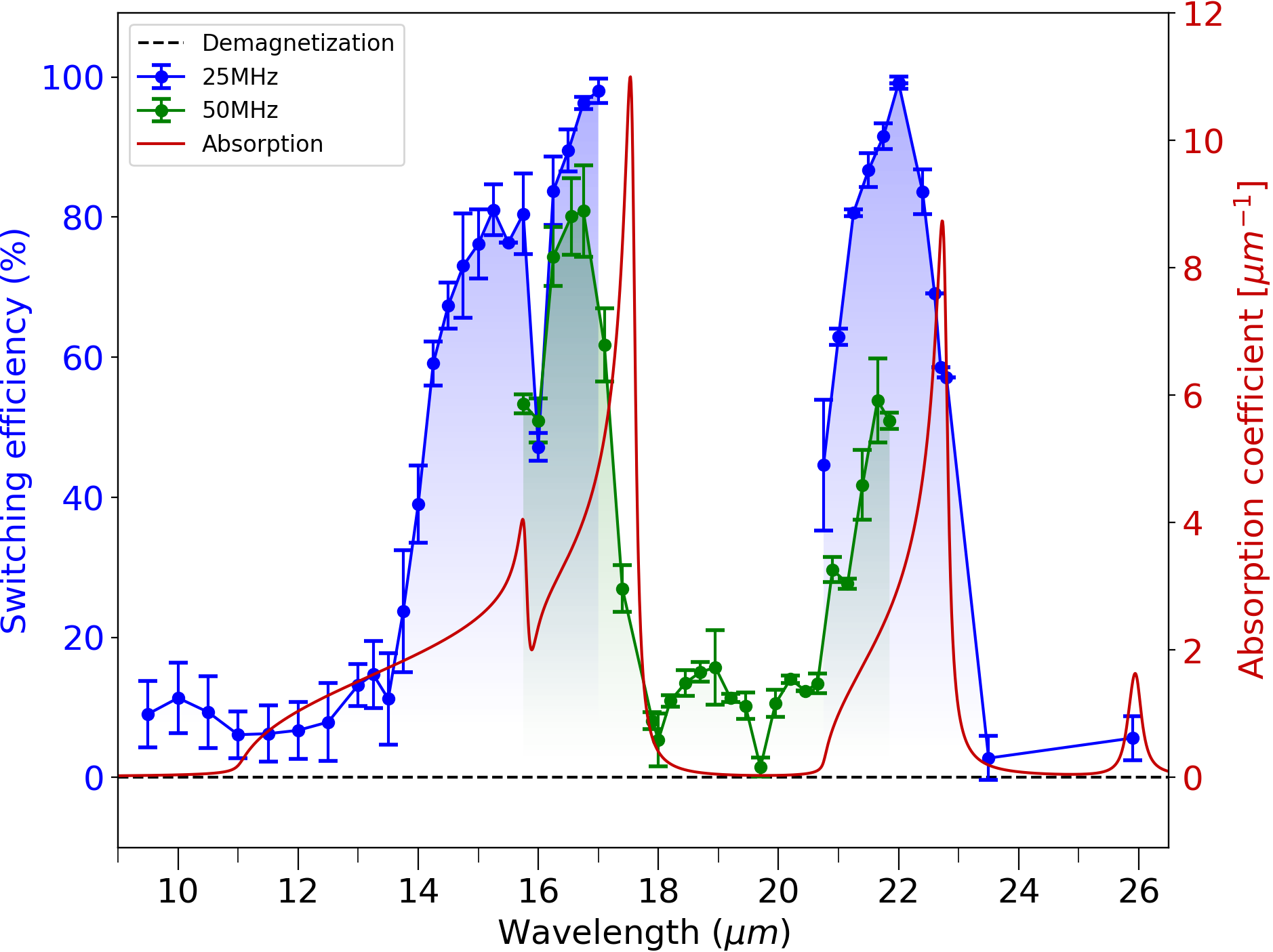}
    \caption{\textbf{Resonant helicity-dependent switching of magnetization}. Spectral dependence of helicity-dependent magnetization switching in a sapphire-mounted GdFeCo sample measured in different experimental runs at room temperature. The blue and green traces show measurements using 25 MHz and 50 MHz laser repetition rates, respectively. While the 25 MHz data (blue) captures switching at the spectral edges, the higher-power 50 MHz measurement (green) was required to enable switching behavior in the mid-range. The red curve indicates the absorption coefficient of the sapphire substrate for reference \cite{schubert_infrared_2000}.}
    \label{fig:fig3}
\end{figure}

\begin{figure}[h!]
    \centering
      \includegraphics[width=\linewidth]{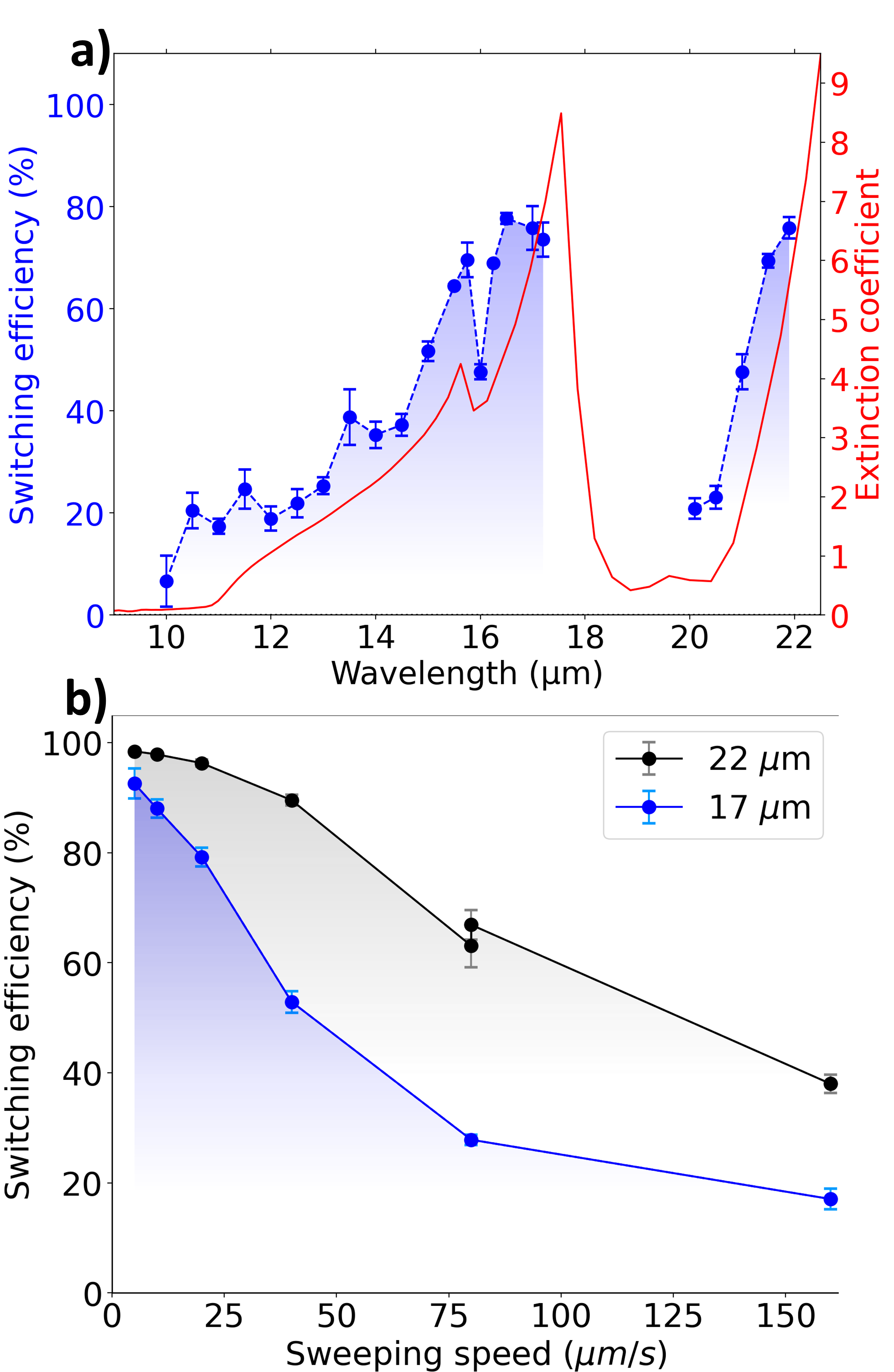}
    \caption{\textbf{Resonant helicity-dependent switching of magnetization}. (a) Spectral dependence of helicity-dependent magnetization switching in a sapphire-mounted GdFeCo sample measured at 400 K extracted from Fig. \ref{fig:sfig_400K}. The blue trace shows the switching efficiency, while the red curve indicates the absorption coefficient of the sapphire substrate for reference \cite{schubert_infrared_2000}. (b) Switching efficiency as a function of sweeping speed at two wavelengths, extracted from Fig. \ref{fig:sfig_speed}.}
    \label{fig:fig4}
\end{figure}

To quantify the switching, we first calculate the difference between a maximum and a minimum of the magnetization state trace. This value is then normalized by the maximum contrast, given by the difference in summed pixel values between completely white and black columns. The resulting ratio expresses the switching quality as a percentage for a given maximum-minimum pair. However, not all extrema are considered when determining the overall switching efficiency of a measurement. For example, in Fig.\ \ref{fig:snormal_track}, the two outermost maxima of the magnetization state trace may arise from either effective switching on the light background or insufficient laser power, which could prevent demagnetization or switching. In such ambiguous cases, the corresponding switching qualities are excluded from the total switching efficiency reported in Figs\ \ref{fig:fig3} and \ref{fig:fig4}.

As will be discussed later, the measured switching efficiency can be influenced by external factors such as the sample temperature and the speed at which the laser spot is swept across the sample. This was achieved by mounting the sample on a motorized translation stage with adjustable speed. When temperature control was required, the standard mount was replaced with a heater, where the sample was attached to a copper plate using silver paste to improve thermal conductivity.

The exemplary magneto-optical images shown in Fig.\ \ref{fig:fig2} reveal a clear dependence of the magnetic switching on the transient grating's wavelength. At $\lambda=$ 10 \um, the IR light caused only demagnetization; switching became apparent at $\lambda=$ 14 \um\ and was nearly perfect at $\lambda=$ 17 \um, resulting in a square-like magnetization state trace. 
As previously noted, the spatial polarization varied from left-handed, through elliptical and linear, to right-handed polarization. When comparing the magnetization state trace with the calculated ellipticity in Fig.\ \ref{fig:fig2} at $\lambda=$ 17 \um, it becomes evident that switching efficiency remains high despite a significant change in ellipticity. This suggests that elliptically-polarized light can still drive effective switching of magnetization at this wavelength. Similar behavior was observed in several other images shown in Fig.\ \ref{fig:sfig1} (corresponding to wavelengths in the vicinity of $\lambda=$ 15.25 \um\ and 21.75 \um), clearly demonstrating that even substantial deviations away from circular polarization can have a negligible effect on switching at these wavelengths.

In contrast, Fig.\ \ref{fig:fig2} at $\lambda =$ 14 \um\ shows that a slight change in circular polarization resulted in a pronounced reduction in switching quality. Thus, farther from resonance, well-defined magnetization reversal is achieved only with near-perfect circular polarization.

By extracting the switching efficiencies from images in Supplementary Figs. \ref{fig:sfig1} and \ref{fig:sfig3}, we constructed the traces shown in Fig.\ \ref{fig:fig3}.
This figure presents the spectral dependence of the switching efficiency derived from multiple measurements of the sapphire-mounted GdFeCo. Importantly, the two experimental traces reasonably align with the substrate's absorption coefficient (shown in red), although slight deviations were observed. This finding underscores the relationship between switching efficiency and the absorption via the excitation of TO phonons within the substrate. In particular, the absorption coefficient of sapphire directly links (in the spectral range of interest \cite{schubert_infrared_2000}) to four distinct IR-active $E_u$ phonon modes with a dipole moment oscillating perpendicular to the c-axis, with TO modes at wavelengths $\lambda =$ 15.8 \um, 17.6 \um, 22.8 \um\, and 26.0 \um.

The blue trace in Fig.\ \ref{fig:fig3} represents the switching efficiency induced by micropulses from the FEL at 25 MHz. It demonstrated demagnetization and no switching beyond $\lambda =$ 23 \um, despite a small substrate absorption peak around $\lambda =$ 26 \um. Moreover, the laser power was insufficient to induce any observable magnetic response between $\lambda =$ 17 \um\ and 20.75 \um. To overcome this power limitation, the micropulse repetition rate was increased to 50 MHz, and the spectral range was remeasured, producing the green trace.
 Notably, the 25 and 50 MHz traces exhibit differing maximum switching efficiencies. The higher repetition rate increases the total energy deposited on the sample, leading to higher temperatures that reduce magnetization and thereby lower the maximum switching efficiency. Nevertheless, the trace reveals a pronounced decrease in switching efficiency as the sapphire's absorption vanishes, closely following the absorption spectrum.

To prove the latter point, Fig.\ \ref{fig:fig4}a shows the influence of increased sample temperature on the switching efficiency, where measurements were performed at 400 K under exposure to 25 MHz laser light. The data confirm that higher temperatures result in lower switching efficiency compared to colder conditions, as observed in Fig. \ref{fig:fig3}. Nevertheless, while overall switching efficiency is reduced at 400 K, greater sensitivity to switching at shorter wavelengths was observed. This suggests that at room temperature, the light did not provide sufficient stimulus to switch the magnetization; however, at elevated temperatures, the added thermal energy facilitated or even enabled magnetization reversal.

Fig. \ref{fig:fig4}b presents the switching efficiency as a function of the sample's sweeping speed at two wavelengths.
It shows that a lower sweeping speed resulted in higher switching efficiency and vice versa. Since the sample is exposed to more micropulses at lower speeds of sweeping, one might expect the sample's temperature to increase, causing the switching efficiency to decrease. However, this effect is insignificant since the time between macropulses is 100 ms, and the heat dissipation occurs on a faster timescale. Instead, the slower sweeping speed allowed more pulses to interact with the same sample area, aiding the switching. 
Adjusting the sweeping speed thus provides a tool to improve switching sensitivity, similar to increasing the temperature, but without compromising the maximum switching efficiency.

Our findings thus confirm substrate-mediated helicity-dependent magnetization switching as a robust mechanism in the infrared spectral regime. Transient polarization grating spectroscopy enabled a broader range of excitation wavelengths and clearly revealed the switching process’s sensitivity to the ellipticity of the incident light.

Despite this overall robustness, variations in switching quality were observed across some of the images in Figs. \ref{fig:sfig1} and \ref{fig:sfig3}. Under ideal conditions, the maxima and minima of the magnetization state trace reach consistent y-values across the sample, as observed in most measurements. However, slight misalignments likely introduced irregularities, leading to deviations from the expected uniformity, as seen in Fig. \ref{fig:sfig1} at $\lambda\ =$ 14.75 and 21.25 $\micro m$ and Fig. \ref{fig:sother_track}. These irregularities required adjustments in the data characterization, specifically in selecting extrema for switching efficiency calculations. 

As described in connection with Fig. \ref{fig:snormal_track}, the outermost extrema were discarded in the standard data processing; accordingly, for Fig. \ref{fig:sother_track}, the leftmost valley and the rightmost peak were excluded. However, when irregularities were observed, an additional data processing step was applied to all measurements, in which only the two largest deltas between the extrema were considered when determining the switching efficiencies. While this method avoided averaging over inconsistent switching behaviors, it also reduced the number of available data points, leading to artificially small error bars and a potential underestimation of measurement uncertainties. 

Despite these experimental constraints, the transient grating setup has provided key advantages. This method is fully compatible with a broad infrared wavelength range generated by the free-electron laser. Additionally, eliminating absorption losses associated with quarter-waveplates or their equivalents \cite{wojdyla_achromatic_2011} increased laser fluence. Transient grating spectroscopy also enabled simultaneous measurement of both helicities, elliptical, and linear polarizations, allowing for faster and more comprehensive analysis of the polarization dependence of switching.

These experimental advantages have allowed us to demonstrate the efficiency and robustness of this novel substrate-mediated switching mechanism. This process enabled deterministic writing and erasing of magnetic domains, with the potential for fast and energy-efficient control. Moreover, its substrate-mediated nature suggests it could be universally applicable to different magnetic layers. Future work will explore the extent of this mechanism across multiple substrates and magnetic materials, clarifying its constraints and capabilities in helicity-dependent magnetization control.\newline

\bibliographystyle{unsrt-nonote}
\bibliography{references.bib}

\begin{thebibliography}{10}

\bibitem{masanet_recalibrating_2020}
Eric Masanet, Arman Shehabi, Nuoa Lei, Sarah Smith, and Jonathan Koomey.
\newblock Recalibrating global data center energy-use estimates.
\newblock {\em Science}, 367(6481):984--986, February 2020.

\bibitem{seagate_technology_llc_why_nodate}
Seagate~Technology LLC.
\newblock Why {HDDs} {Dominate} {Hyperscale} {Cloud} {Architecture}.
\newblock https://www.seagate.com/blog/why-hdds-dominate-hyperscale-cloud-architecture/,  (accessed 2024-04-22).

\bibitem{stanciu_all-optical_2007}
C.~D. Stanciu, F.~Hansteen, A.~V. Kimel, A.~Kirilyuk, A.~Tsukamoto, A.~Itoh, and Th. Rasing.
\newblock All-{Optical} {Magnetic} {Recording} with {Circularly} {Polarized} {Light}.
\newblock {\em Physical Review Letters}, 99(4):047601, July 2007.

\bibitem{kirilyuk_laser-induced_2013}
Andrei Kirilyuk, Alexey~V. Kimel, and Theo Rasing.
\newblock Laser-induced magnetization dynamics and reversal in ferrimagnetic alloys.
\newblock {\em Reports on Progress in Physics}, 76(2):026501, February 2013.

\bibitem{vahaplar_ultrafast_2009}
K.~Vahaplar, A.~M. Kalashnikova, A.~V. Kimel, D.~Hinzke, U.~Nowak, R.~Chantrell, A.~Tsukamoto, A.~Itoh, A.~Kirilyuk, and Th. Rasing.
\newblock Ultrafast {Path} for {Optical} {Magnetization} {Reversal} via a {Strongly} {Nonequilibrium} {State}.
\newblock {\em Physical Review Letters}, 103(11):117201, September 2009.

\bibitem{radu_transient_2011}
I.~Radu, K.~Vahaplar, C.~Stamm, T.~Kachel, N.~Pontius, H.~A. Dürr, T.~A. Ostler, J.~Barker, R.~F.~L. Evans, R.~W. Chantrell, A.~Tsukamoto, A.~Itoh, A.~Kirilyuk, Th. Rasing, and A.~V. Kimel.
\newblock Transient ferromagnetic-like state mediating ultrafast reversal of antiferromagnetically coupled spins.
\newblock {\em Nature}, 472(7342):205--208, April 2011.

\bibitem{ostler_ultrafast_2012}
T.~A. Ostler, J.~Barker, R.~F.~L. Evans, R.~W. Chantrell, U.~Atxitia, O.~Chubykalo-Fesenko, S.~El~Moussaoui, L.~Le~Guyader, E.~Mengotti, L.~J. Heyderman, F.~Nolting, A.~Tsukamoto, A.~Itoh, D.~Afanasiev, B.~A. Ivanov, A.~M. Kalashnikova, K.~Vahaplar, J.~Mentink, A.~Kirilyuk, Th. Rasing, and A.~V. Kimel.
\newblock Ultrafast heating as a sufficient stimulus for magnetization reversal in a ferrimagnet.
\newblock {\em Nature Communications}, 3(1):666, February 2012.

\bibitem{mentink_ultrafast_2012}
J.~H. Mentink, J.~Hellsvik, D.~V. Afanasiev, B.~A. Ivanov, A.~Kirilyuk, A.~V. Kimel, O.~Eriksson, M.~I. Katsnelson, and Th. Rasing.
\newblock Ultrafast {Spin} {Dynamics} in {Multisublattice} {Magnets}.
\newblock {\em Physical Review Letters}, 108(5):057202, January 2012.

\bibitem{davies_helicity-independent_2022}
C.~S. Davies, J.~H. Mentink, A.~V. Kimel, Th. Rasing, and A.~Kirilyuk.
\newblock Helicity-independent all-optical switching of magnetization in ferrimagnetic alloys.
\newblock {\em Journal of Magnetism and Magnetic Materials}, 563:169851, December 2022.

\bibitem{khorsand_role_2012}
A.~R. Khorsand, M.~Savoini, A.~Kirilyuk, A.~V. Kimel, A.~Tsukamoto, A.~Itoh, and Th. Rasing.
\newblock Role of {Magnetic} {Circular} {Dichroism} in {All}-{Optical} {Magnetic} {Recording}.
\newblock {\em Physical Review Letters}, 108(12):127205, March 2012.

\bibitem{mangin_engineered_2014}
S.~Mangin, M.~Gottwald, C.-H. Lambert, D.~Steil, V.~Uhlíř, L.~Pang, M.~Hehn, S.~Alebrand, M.~Cinchetti, G.~Malinowski, Y.~Fainman, M.~Aeschlimann, and E.~E. Fullerton.
\newblock Engineered materials for all-optical helicity-dependent magnetic switching.
\newblock {\em Nature Materials}, 13(3):286--292, March 2014.

\bibitem{lambert_all-optical_2014}
C-H. Lambert, S.~Mangin, B.~S. D. Ch.~S. Varaprasad, Y.~K. Takahashi, M.~Hehn, M.~Cinchetti, G.~Malinowski, K.~Hono, Y.~Fainman, M.~Aeschlimann, and E.~E. Fullerton.
\newblock All-optical control of ferromagnetic thin films and nanostructures.
\newblock {\em Science}, 345(6202):1337--1340, September 2014.

\bibitem{medapalli_multiscale_2017}
R.~Medapalli, D.~Afanasiev, D.~K. Kim, Y.~Quessab, S.~Manna, S.~A. Montoya, A.~Kirilyuk, Th. Rasing, A.~V. Kimel, and E.~E. Fullerton.
\newblock Multiscale dynamics of helicity-dependent all-optical magnetization reversal in ferromagnetic {Co}/{Pt} multilayers.
\newblock {\em Physical Review B}, 96(22):224421, December 2017.

\bibitem{kichin_multiple-_2019}
G.~Kichin, M.~Hehn, J.~Gorchon, G.~Malinowski, J.~Hohlfeld, and S.~Mangin.
\newblock From {Multiple}- to {Single}-{Pulse} {All}-{Optical} {Helicity}-{Dependent} {Switching} in {Ferromagnetic} {Co}/{Pt} {Multilayers}.
\newblock {\em Physical Review Applied}, 12(2):024019, August 2019.

\bibitem{yamada_efficient_2022}
Kihiro~T. Yamada, Alexey~V. Kimel, Kiran~Horabail Prabhakara, Sergiu Ruta, Tian Li, Fuyuki Ando, Sergey Semin, Teruo Ono, Andrei Kirilyuk, and Theo Rasing.
\newblock Efficient {All}-{Optical} {Helicity} {Dependent} {Switching} of {Spins} in a {Pt}/{Co}/{Pt} {Film} by a {Dual}-{Pulse} {Excitation}.
\newblock {\em Frontiers in Nanotechnology}, 4, February 2022.

\bibitem{stupakiewicz_ultrafast_2017}
A.~Stupakiewicz, K.~Szerenos, D.~Afanasiev, A.~Kirilyuk, and A.~V. Kimel.
\newblock Ultrafast nonthermal photo-magnetic recording in a transparent medium.
\newblock {\em Nature}, 542(7639):71--74, February 2017.

\bibitem{stupakiewicz_ultrafast_2021}
A.~Stupakiewicz, C.~S. Davies, K.~Szerenos, D.~Afanasiev, K.~S. Rabinovich, A.~V. Boris, A.~Caviglia, A.~V. Kimel, and A.~Kirilyuk.
\newblock Ultrafast phononic switching of magnetization.
\newblock {\em Nature Physics}, 17(4):489--492, 2021.

\bibitem{davies_phononic_2024}
C.~S. Davies, F.~G.~N. Fennema, A.~Tsukamoto, I.~Razdolski, A.~V. Kimel, and A.~Kirilyuk.
\newblock Phononic switching of magnetization by the ultrafast {Barnett} effect.
\newblock {\em Nature}, 628:540--544, April 2024.

\bibitem{dornes_ultrafast_2019}
C.~Dornes, Y.~Acremann, M.~Savoini, M.~Kubli, M.~J. Neugebauer, E.~Abreu, L.~Huber, G.~Lantz, C.~a.~F. Vaz, H.~Lemke, E.~M. Bothschafter, M.~Porer, V.~Esposito, L.~Rettig, M.~Buzzi, A.~Alberca, Y.~W. Windsor, P.~Beaud, U.~Staub, Diling Zhu, Sanghoon Song, J.~M. Glownia, and S.~L. Johnson.
\newblock The ultrafast {Einstein}–de {Haas} effect.
\newblock {\em Nature}, 565(7738):209--212, January 2019.

\bibitem{tauchert_polarized_2022}
S.~R. Tauchert, M.~Volkov, D.~Ehberger, D.~Kazenwadel, M.~Evers, H.~Lange, A.~Donges, A.~Book, W.~Kreuzpaintner, U.~Nowak, and P.~Baum.
\newblock Polarized phonons carry angular momentum in ultrafast demagnetization.
\newblock {\em Nature}, 602(7895):73--77, February 2022.

\bibitem{juraschek_phono-magnetic_2020}
Dominik~M. Juraschek, Prineha Narang, and Nicola~A. Spaldin.
\newblock Phono-magnetic analogs to opto-magnetic effects.
\newblock {\em Physical Review Research}, 2(4):043035, October 2020.

\bibitem{basini_terahertz_2024}
M.~Basini, M.~Pancaldi, B.~Wehinger, M.~Udina, V.~Unikandanunni, T.~Tadano, M.~C. Hoffmann, A.~V. Balatsky, and S.~Bonetti.
\newblock Terahertz electric-field-driven dynamical multiferroicity in {SrTiO3}.
\newblock {\em Nature}, 628(8008):534--539, April 2024.

\bibitem{luo_large_2023}
Jiaming Luo, Tong Lin, Junjie Zhang, Xiaotong Chen, Elizabeth~R. Blackert, Rui Xu, Boris~I. Yakobson, and Hanyu Zhu.
\newblock Large effective magnetic fields from chiral phonons in rare-earth halides.
\newblock {\em Science}, 382(6671):698--702, November 2023.

\bibitem{stanciu_ultrafast_2006}
C.~D. Stanciu, A.~V. Kimel, F.~Hansteen, A.~Tsukamoto, A.~Itoh, A.~Kirilyuk, and Th. Rasing.
\newblock Ultrafast spin dynamics across compensation points in ferrimagnetic {GdFeCo}: {The} role of angular momentum compensation.
\newblock {\em Physical Review B}, 73(22):220402, June 2006.

\bibitem{oepts_free-electron-laser_1995}
D.~Oepts, A.~F.~G. van~der Meer, and P.~W. van Amersfoort.
\newblock The {Free}-{Electron}-{Laser} user facility {FELIX}.
\newblock {\em Infrared Physics \& Technology}, 36(1):297--308, January 1995.

\bibitem{knippels_fel_1998}
G.~M.~H. Knippels and A.~F.~G. van~der Meer.
\newblock {FEL} diagnostics and user control.
\newblock {\em Nuclear Instruments and Methods in Physics Research Section B: Beam Interactions with Materials and Atoms}, 144(1):32--39, September 1998.

\bibitem{janusonis_magnetoelastic_2017}
Julius Janusonis.
\newblock {\em Magnetoelastic waves in thin ferromagnetic films}.
\newblock University of Groningen, [Groningen], 2017.

\bibitem{davies_towards_2018}
C.~S. Davies, J.~Janušonis, A.~V. Kimel, A.~Kirilyuk, A.~Tsukamoto, Th. Rasing, and R.~I. Tobey.
\newblock Towards massively parallelized all-optical magnetic recording.
\newblock {\em Journal of Applied Physics}, 123(21):213904, June 2018.

\bibitem{schubert_infrared_2000}
M.~Schubert, T.~E. Tiwald, and C.~M. Herzinger.
\newblock Infrared dielectric anisotropy and phonon modes of sapphire.
\newblock {\em Physical Review B}, 61(12):8187--8201, March 2000.

\bibitem{wojdyla_achromatic_2011}
A.~Wojdyla and G.~Gallot.
\newblock Achromatic polarizing elements for pulsed {THz} waves.
\newblock In {\em 2011 {International} {Conference} on {Infrared}, {Millimeter}, and {Terahertz} {Waves}}, pages 1--2, October 2011.
\newblock ISSN: 2162-2035.

\end{thebibliography}
\section*{ACKNOWLEDGEMENTS}
We thank all technical staff at HFML-FELIX for their technical support. We gratefully acknowledge the Nederlandse Organisatie voor Wetenschappelijk Onderzoek (NWO-I) for their financial contribution, including the support of HFML-FELIX. C.S.D. acknowledges support from the European Research Council ERC Grant Agreement No. 101115234 (HandShake), and A.K. acknowledges support from the European Research Council ERC Grant Agreement No. 101141740 (INTERPHON).

\section*{AUTHOR DECLARATIONS}
\subsection*{Conflict of Interest}
The authors have no conflicts to disclose.

\section*{DATA AVAILABILITY}
The data that support the findings of this study are available from the corresponding author upon reasonable request.

\newpage
\clearpage
\onecolumngrid


\section*{Supplementary Information}
\renewcommand{\thefigure}{S\arabic{figure}} 
\setcounter{figure}{0} 

\subsection*{Supplementary Note 1: Background-corrected Images Supporting Switching Efficiency Analysis }
The following magneto-optical images show background-corrected measurements underlying the switching efficiency analysis presented in Fig. \ref{fig:fig3}. 

\begin{figure}[h]
    \centering
    \includegraphics[width=1\linewidth, height=22cm, keepaspectratio=false]{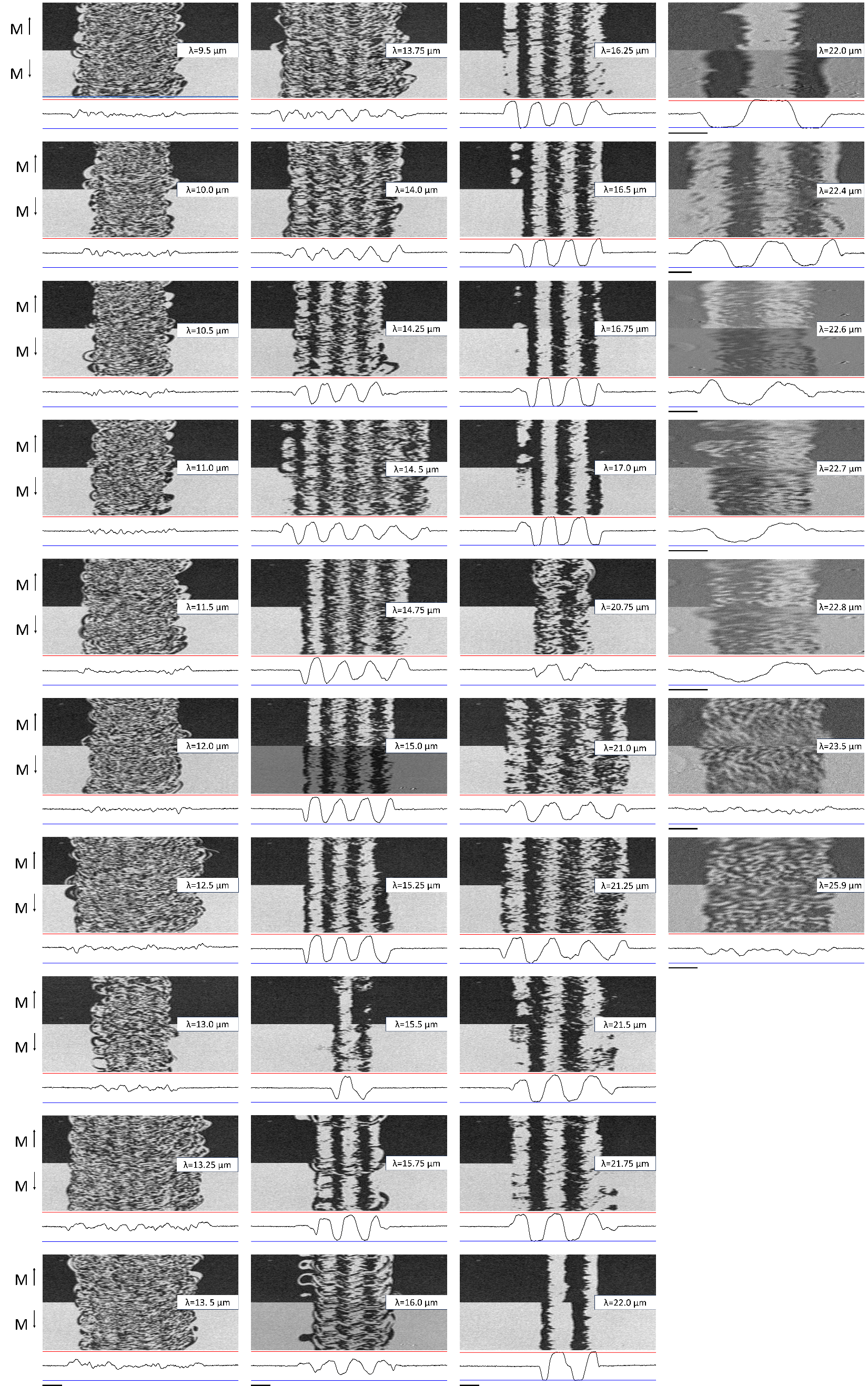}
    \caption{Processed magneto-optical images depicting the magnetization of the sapphire-mounted  GdFeCo sample after irradiation with IR pulses with different central wavelengths, as indicated in each image. The free-electron laser operated at a repetition rate of 25 MHz, with the sample maintained at 300 K, and the sweeping speed was 20 \um/s.  The scale bars (corresponding to 30 \um) are common for all images. For the rightmost column, individual bars are shown due to the higher magnification. The extracted switching efficiencies are plotted as the blue trace in Fig. \ref{fig:fig3}.}
    \label{fig:sfig1}
\end{figure}

\begin{figure}[h]
    \centering  
    \includegraphics[width=1\linewidth]{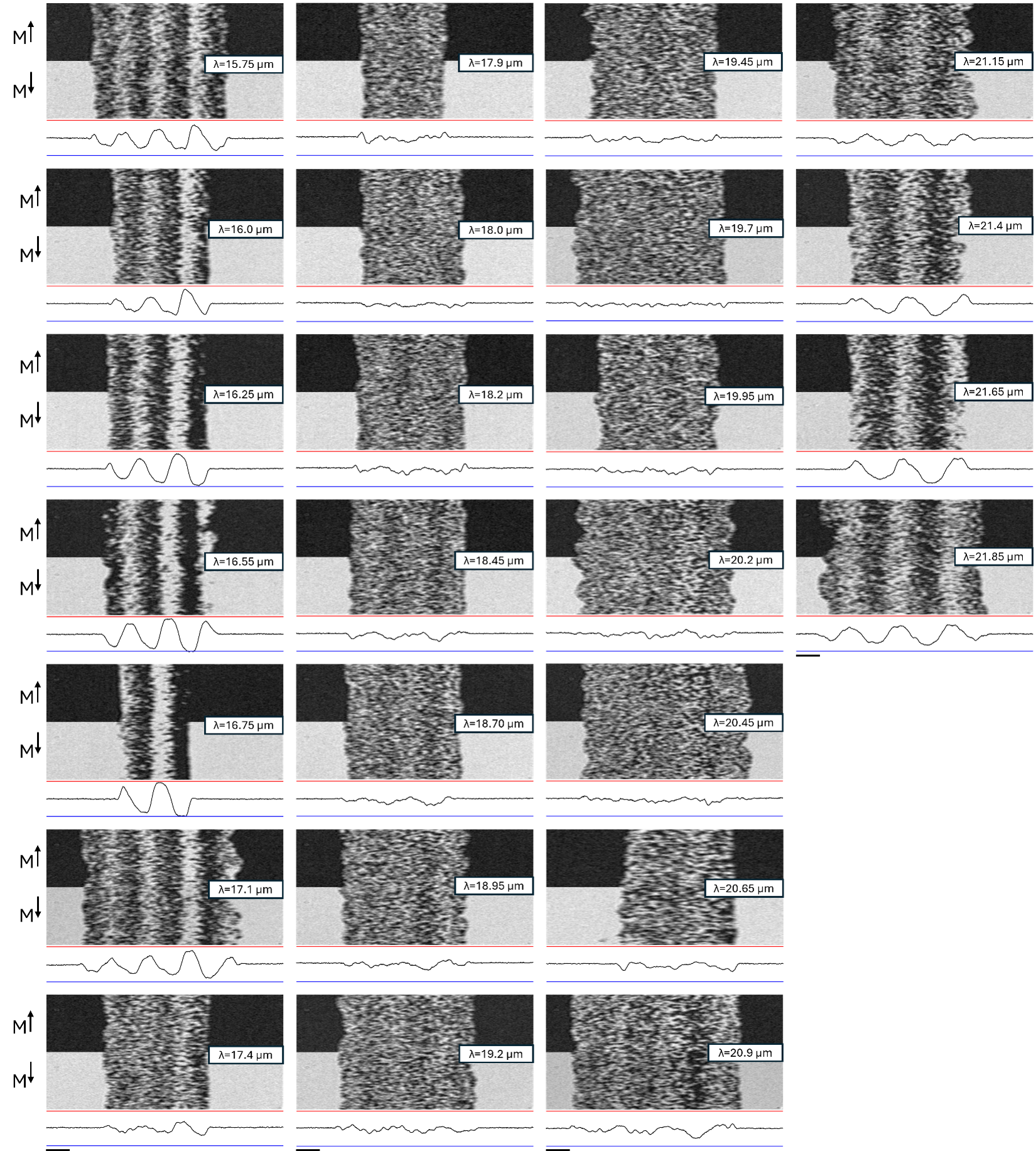}
    \caption{Processed magneto-optical images depicting the magnetization of the sapphire-mounted  GdFeCo sample after irradiation with IR pulses with different central wavelengths, as indicated in each image. The free-electron laser operated at a repetition rate of 50 MHz, with the sample maintained at 300 K, and the sweeping speed was 20 \um/s. The scale bars (corresponding to 30 \um) are common to all images. The extracted switching efficiencies are plotted as the green trace in Fig. \ref{fig:fig3}.}
    \label{fig:sfig3}
\end{figure}

\FloatBarrier

\subsection*{Supplementary Note 2: Extra Example Images Showing Imperfect Alignment  }
\begin{figure}[h]
    \centering  
    \includegraphics[width=0.5\linewidth]{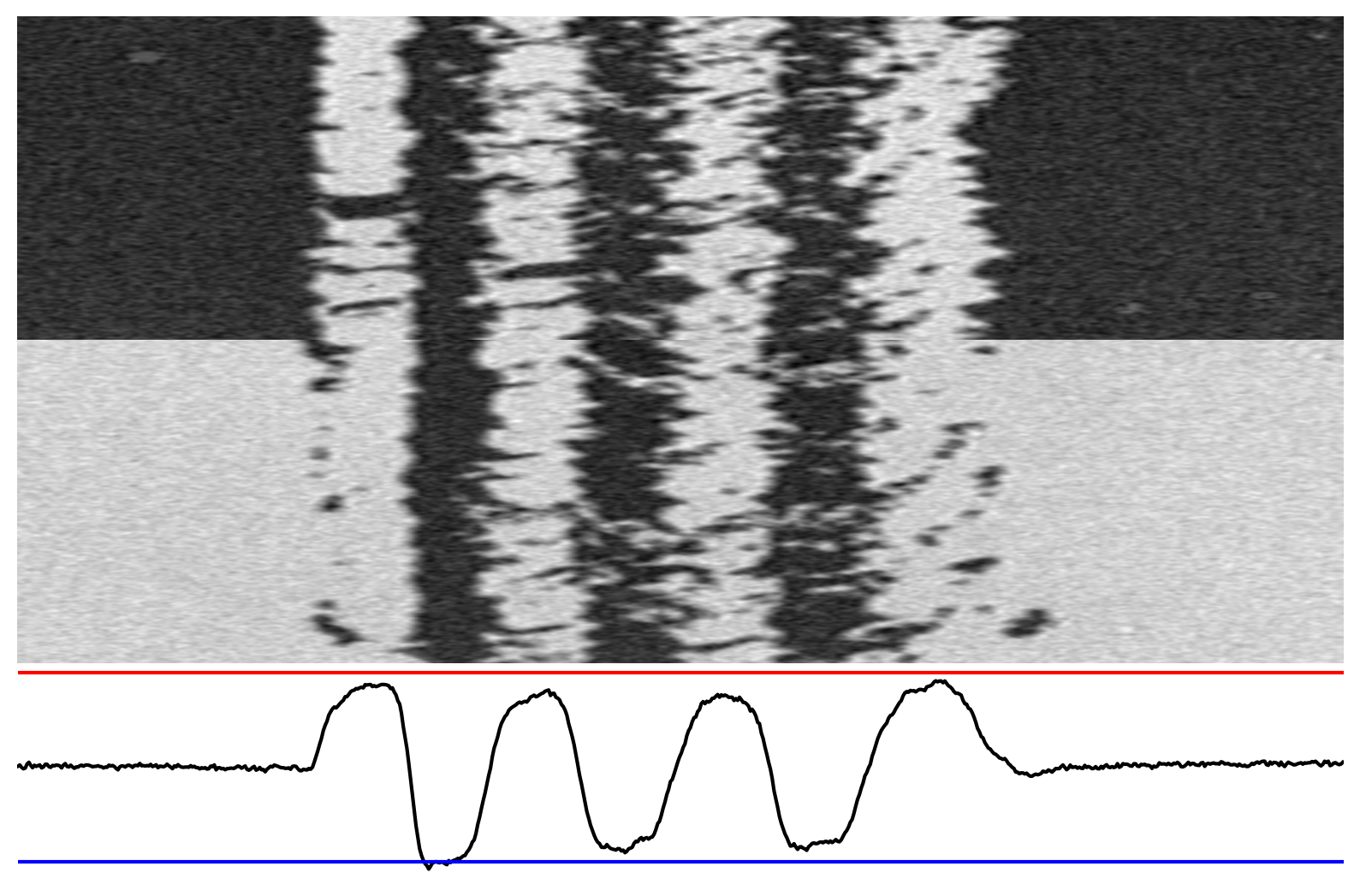}
    \caption{An image from Fig. \ref{fig:sfig1} measured at $\lambda = 21.75\ \micro m$ showing a normal track where the magnetization state trace shows consistent y-values of the extrema. }
    \label{fig:snormal_track}
\end{figure}

\begin{figure}[h]
    \centering  
    \includegraphics[width=0.5\linewidth]{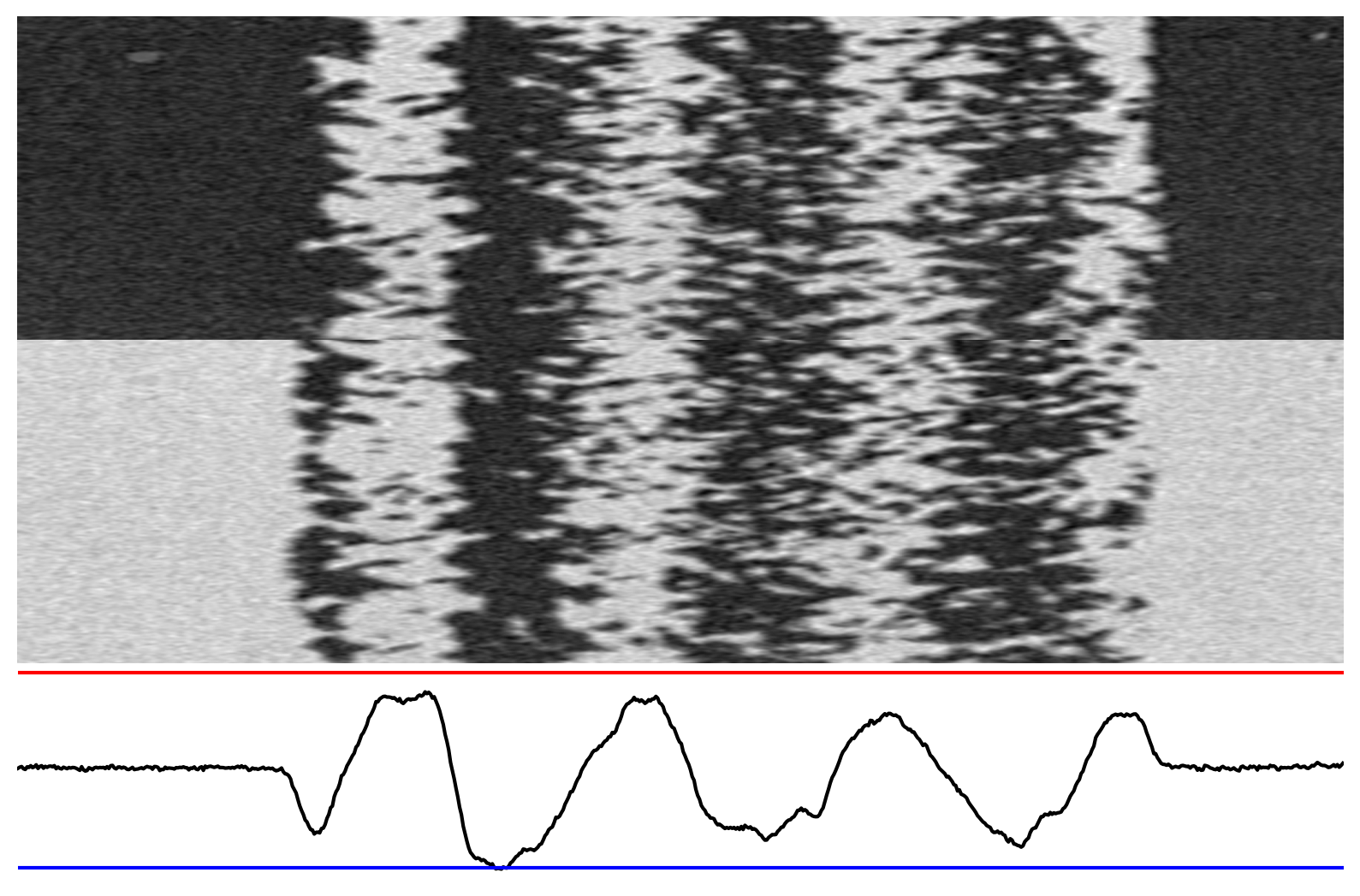}
    \caption{An image from Fig. \ref{fig:sfig1} measured at $\lambda = 14.75\ \micro m$ showing a changing switching efficiency across the sample where the magnetization state trace shows inconsistent y-values of the extrema.}
    \label{fig:sother_track}
\end{figure}

The following magneto-optical images show background-corrected measurements and its associated magnetization state trace. Fig. \ref{fig:snormal_track} shows consistent values for the extrema of the black trace. Oppositely, Fig. \ref{fig:sother_track} shows a damped oscillation in amplitude of the magnetization state trace. These irregularities required adjustments in the data characterization, and only the highest maximum and lowest minimum are considered for the switching efficiency calculation.  

\newpage
\subsection*{Supplementary Note 3: Background-corrected Images Supporting Switching Efficiency Analysis}
The following magneto-optical images show background-corrected measurements underlying the switching efficiency analysis presented in Fig. \ref{fig:fig4}. 

\begin{figure}[h]
    \centering  
    \includegraphics[width=1\linewidth]{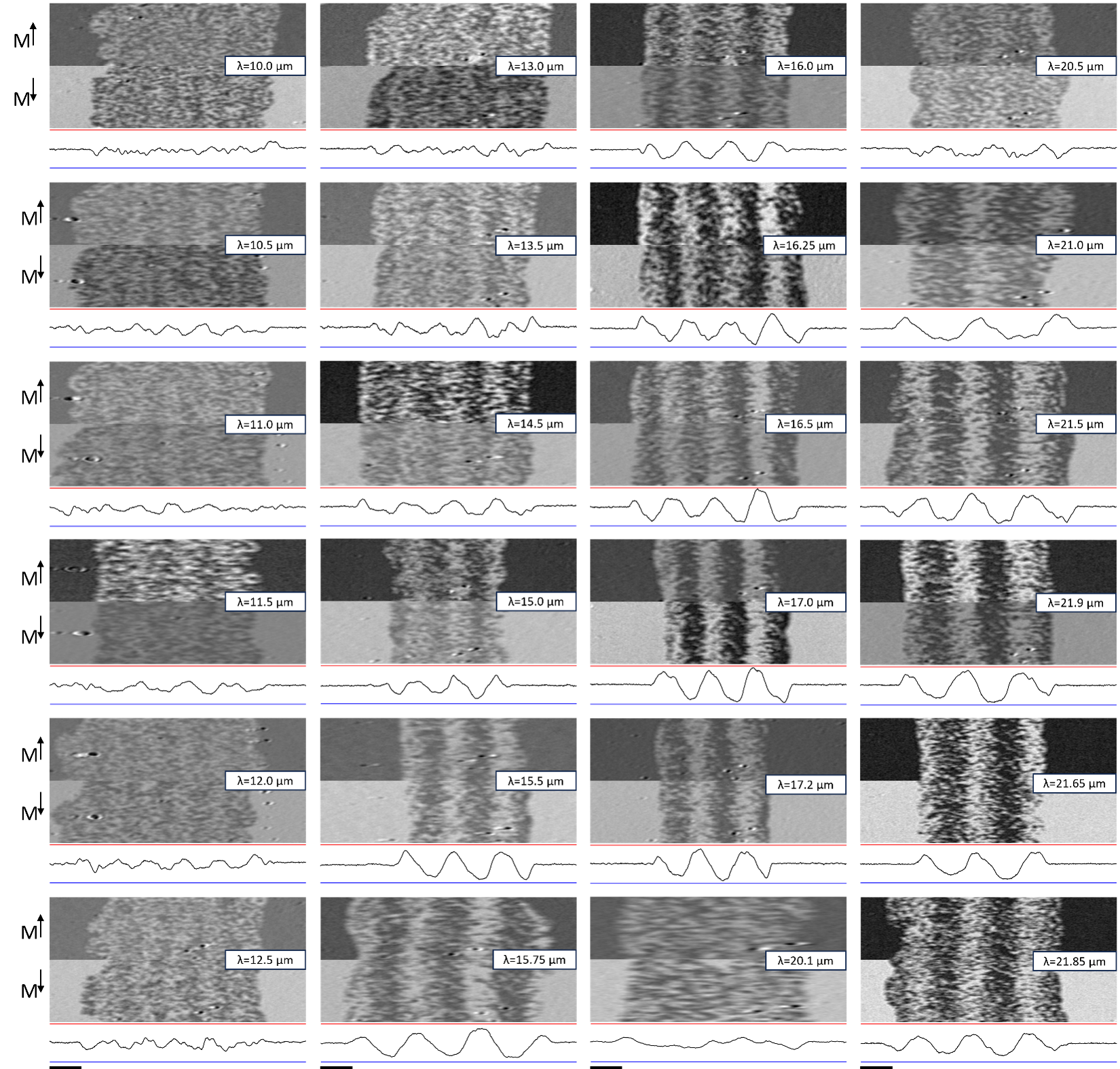}
    \caption{Processed magneto-optical images depicting the magnetization of the sapphire-mounted  GdFeCo sample after irradiation with IR pulses with different central wavelengths, as indicated in each image. The free-electron laser operated at a repetition rate of 25 MHz, with the sample maintained at 400 K, and the sweeping speed was 20 \um/s. The scale bars (corresponding to 50 \um) are common to all images. The extracted switching efficiencies are plotted as the blue trace in Fig. \ref{fig:fig4}a.}
    \label{fig:sfig_400K}
\end{figure}

\begin{figure}[h]
    \centering  
    \includegraphics[width=0.6\linewidth]{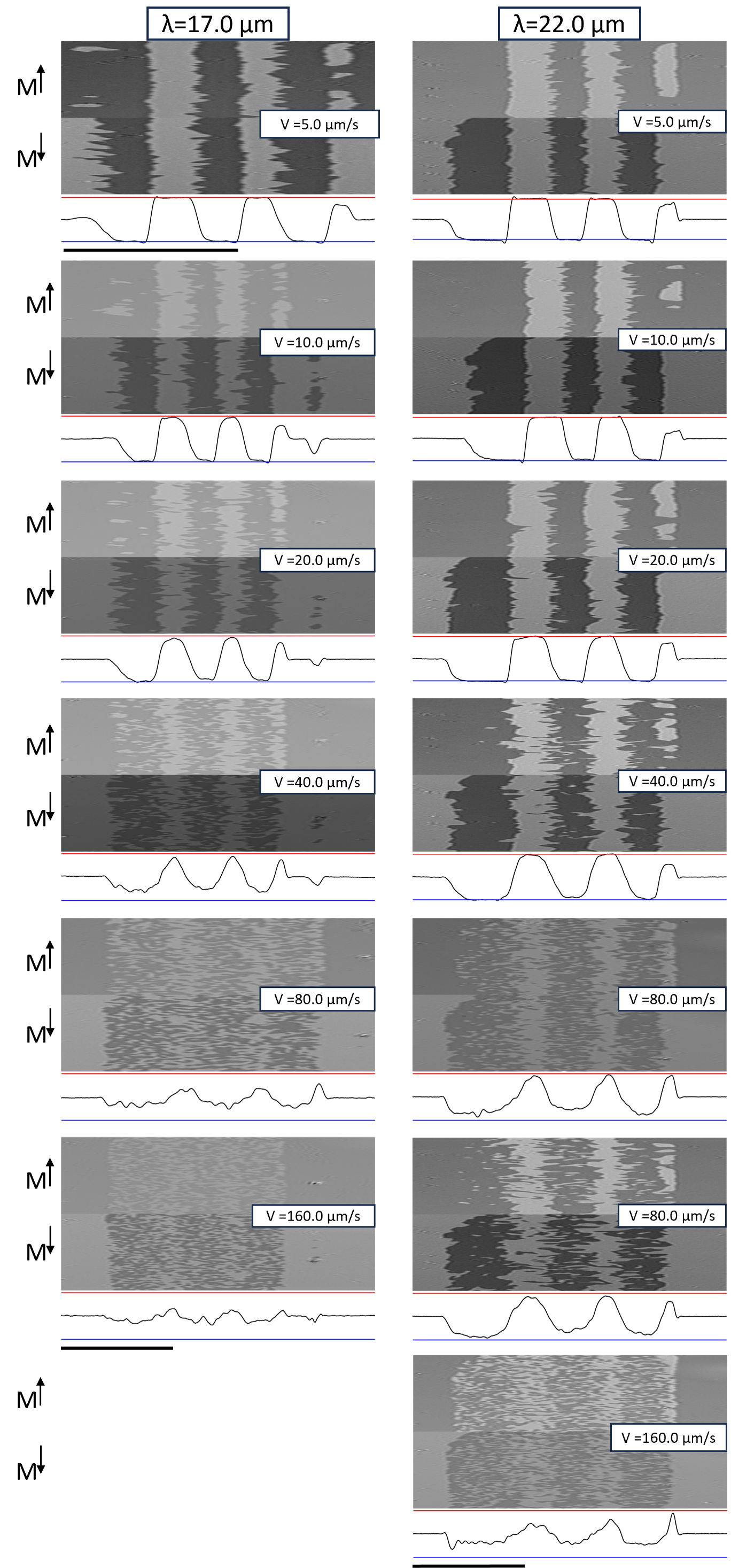}
    \caption{Processed magneto-optical images depicting the magnetization of the sapphire-mounted  GdFeCo sample after irradiation with IR pulses with different sweeping speeds and central wavelengths, as indicated in each image. The free-electron laser operated at a repetition rate of 25 MHz, with the sample maintained at 300 K. The scale bars (corresponding to 50 \um) are common to all images, except in the upper-left panel, where an individual bar is shown due to the higher magnification. The extracted switching efficiencies are plotted in Fig. \ref{fig:fig4}b.}
    \label{fig:sfig_speed}
\end{figure}
\FloatBarrier

\subsection*{Supplementary Note 4: Jones Calculus}

In order to calculate the helicity shown in Fig. \ref{fig:fig2}, Jones matrices are used. Our light, which is created by the interference of two pulses under a variable angle, is modeled by the combination of horizontal and vertical light beams that travel through left- and right-circular polarizers. 

\bigskip
\begin{flushleft}The Jones matrix for a right-circular polarizer is given as: \end{flushleft}

\begin{equation*}
    \frac{1}{2} \begin{pmatrix}1 & i \\-i & 1\end{pmatrix},
\end{equation*}
\begin{flushleft}Which can be used to combine two orthogonally polarized light pulses to form a right-circularly polarized light pulse: \end{flushleft} 
\begin{equation}
    \frac{1}{2} \begin{pmatrix}1 & i \\-i & 1\end{pmatrix} \cdot \left(E_H\begin{pmatrix}1\\0\end{pmatrix} + E_V \begin{pmatrix}0\\1\end{pmatrix}\right)= \frac{E_{RCP}}{\sqrt{2}} \begin{pmatrix}1\\-i\end{pmatrix} .
\label{eq:rcp}
\end{equation}

\begin{flushleft}Where the electric field can be written as: \end{flushleft}
\begin{equation*}
     E_\alpha=A_\alpha e^{i\varphi_\alpha}
\end{equation*}

\begin{flushleft}By solving Equation \ref{eq:rcp}, along with a similar approach for a left-circular polarizer, the amplitudes and phases of the helicities can be determined, as shown in Equation \ref{eq:ampandpahse}. \end{flushleft}

\begin{equation}
    \begin{aligned}
        A_R,A_L &=\sqrt{A_H^2+A_V^2 + 2A_HA_V\cos(-\varphi_H+\varphi_V\pm\tfrac{\pi}{2})} \\
        \varphi_R,\varphi_L &= \arctan\left(\frac{A_V\sin(-\varphi_H+\varphi_V\pm\frac{\pi}{2})}{A_H+A_V \cos(-\varphi_H+\varphi_V\pm\frac{\pi}{2}}\right)+\varphi_V \\
        \text{Where,}\\
        \varphi_{H,V}&=\frac{2\pi x}{\lambda}\sin(\pm \alpha_{AOI})\\
         \text{And: } \ & \alpha_{AOI} \ \text{is the angle of incidence.}
    \end{aligned}
\label{eq:ampandpahse}
\end{equation}
On the sample, left- and right-circularly polarized light overlap, with a phase shift between the two pulses that depends on the light wavelength. Next, by subtracting the amplitudes of both circular polarizations, the net circular amplitude is obtained, as given by:

\begin{equation*}
    A_{circular} = A_R-A_L
\end{equation*}

\begin{flushleft}This is plotted as the orange traces in Fig. \ref{fig:fig2}, revealing a spatially varying polarization state.\end{flushleft}

\end{document}